\documentclass[iop]{emulateapj}
\usepackage{enumitem}
\usepackage{graphicx,url}
\usepackage{color}
\usepackage[backref,breaklinks,colorlinks,citecolor=blue]{hyperref}

\def\j1752{XTE~J1752--223}
\def\gx339{GX~339--4}

\def\rxte{{\it RXTE}}
\def\xmm{{\it XMM-Newton}}
\def\suzaku{{\it Suzaku}}
\def\swift{{\it Swift}}

\def\maxi{{\it MAXI}}

\def\xillverCp{{\tt xillverCp}}
\def\nthComp{{\tt nthComp}}
\def\relxill{{\tt relxill}}
\def\relxilllp{{\tt relxilllp}}
\def\relxillCp{{\tt relxillCp}}
\def\relxilllpCp{{\tt relxilllpCp}}
\def\simplcut{{\tt simplcut}}

\def\pcacorr{{\sc pcacorr}}
\def\hexBcorr{{\sc hexBcorr}}
\def\crabcorr{{\tt crabcorr}}
\def\msun{$M_{\odot}$}

\shorttitle{Reflection Spectroscopy of \j1752}
\shortauthors{Garc\'{\i}a \& et al.}

\setlist[itemize]{leftmargin=*}

\begin{document}

\title{Reflection Spectroscopy of the Black Hole Binary \j1752\ in 
its Long-Stable Hard State}

\author{Javier~A.~Garc\'ia\altaffilmark{1,2},
  James~F.~Steiner\altaffilmark{3}, 
  Victoria~Grinberg\altaffilmark{4}, 
  Thomas~Dauser\altaffilmark{2},
  Riley~M.~T.~Connors\altaffilmark{1},
  Jeffrey~E.~McClintock\altaffilmark{5},
  Ronald~A.~Remillard\altaffilmark{3},
  J\"orn~Wilms\altaffilmark{2},
  Fiona~A.~Harrison\altaffilmark{1},
  and John~A.~Tomsick\altaffilmark{6}
}

\altaffiltext{1}{Cahill Center for Astronomy and Astrophysics,
  California Institute of Technology, Pasadena, CA 91125, USA;
  javier@caltech.edu}

\altaffiltext{2}{Dr.\ Karl Remeis-Observatory and Erlangen Centre for 
  Astroparticle Physics, Sternwartstr.~7, 96049 Bamberg, Germany}

\altaffiltext{3}{MIT Kavli Institute for Astrophysics and Space  Research,
  MIT, 70 Vassar Street, Cambridge, MA 02139}

\altaffiltext{4}{ESA European Space Research and Technology Centre (ESTEC),
  Keplerlaan 1, 2201 AZ Noordwijk, The Netherlands}

\altaffiltext{5}{Harvard-Smithsonian Center for Astrophysics,
  60 Garden St., Cambridge, MA 02138 USA}

\altaffiltext{6}{Space Sciences Laboratory, 7 Gauss Way, University of
  California, Berkeley, CA 94720-7450, USA}

%

\begin{abstract}

We present a detailed spectral analysis of the Black Hole Binary \j1752\ in the
hard state of its 2009 outburst.  Regular monitoring of this source by \rxte\
provided high signal-to-noise spectra along the
outburst rise and decay.  During one full month this source stalled at
$\sim$30\% of its peak count rate at a constant hardness and intensity.  By
combining all the data in this exceptionally-stable hard state, we obtained an
aggregate PCA spectrum (3--45~keV) with 100~million counts, and a corresponding
HEXTE spectrum (20--140~keV) with 5.8~million counts.
Implementing a version of our reflection code with a
physical model for Comptonization, we obtain tight constraints on important
physical parameters for this system. In particular, the inner accretion disk is
measured very close in, at $R_\mathrm{in}=1.7\pm0.4$~$R_g$. Assuming
$R_\mathrm{in}=R_\mathrm{ISCO}$, we find a relatively high black hole
spin ($a_*=0.92\pm0.06$). Imposing a lamppost geometry, we obtain a
low inclination ($i=35\pm4$~deg), which agrees with the upper limit found in
the radio ($i<49$~deg). However, we note that this model cannot be
statistically distinguished from a non-lamppost model with free emissivity
index, for which the inclination is markedly higher.  Additionally, we find a
relatively cool corona ($57-70$~keV), and large iron abundance ($3.3-3.7$
solar).  We further find that properly accounting for Comptonization of the
reflection emission improves the fit significantly and causes an otherwise low
reflection fraction ($\sim 0.2-0.3$) to increase by an order of magnitude, in
line with geometrical expectations for a lamppost corona.  We compare these
results with similar investigations reported for \gx339\ in its bright
hard state.
\end{abstract}


%
%
%
%
\section{Introduction}\label{sec:intro}

Accreting black holes are unique probes of physics under the conditions of
extreme gravity. Supermassive black holes at the centers of galaxies shape their
hosts through their powerful outflows \citep{fab12c}, and their smaller
cousins, black hole binaries (BHBs), may have played an important role during
the epoch of ionization \citep[e.g.;][]{mad17}. Accretion processes in
supermassive black holes are, however, hard to study because of their long
variability timescale and large distances that result in low fluxes.  Luckily,
timescales of key accretion and ejection processes scale with mass, so that
BHBs can be seen as supermassive black holes on fast-forward, with whole
outburst cycles occurring within mere months or years.

Typical BHBs spend most of their time in quiescence. They start an outburst in
the hard state, when the source spectrum is dominated by hard X-rays in the
form of a power-law component with photon index $\Gamma \sim 1.7$. In this state,
there is little or no contribution from thermal accretion-disk emission. Radio
emission is detected and, for some sources, collimated outflows have been
resolved \citep[e.g.;][]{mir+1999x}. The source brightness increases at almost
constant hardness until the spectrum finally begins to soften both in photon
index and through stronger contribution from the accretion disk in the soft
X-rays: the source enters first the hard-intermediate and then the
soft-intermediate states.  It finally reaches the soft state (also referred to
as the thermal state), when the X-ray continuum is dominated by the accretion
disk emission and the power-law component is steep ($\Gamma \gtrsim 2$). Radio
emission in the soft state is strongly suppressed. In a typical outburst, the
source dims over months in the thermal state, and at lower luminosity returns
through intermediate states back to a hard state in which it fades back into
quiescence \citep[see][for reviews]{mcc06,rem06}.

While the phenomenology itself is rather well described, its physical
underpinning remains a mystery. In particular, the geometry of the emission
region is unclear: the power law is likely produced through thermal Comptonization and
further modified through reflection off the accretion disk, but the origin of
this Comptonizing medium---often referred to as the {\it corona}---whether it is
due to inverse Compton (IC) scattering of disk photons in a hot and static plasma
\citep[e.g.;][]{haa93,dov97,zdz03}, or whether it is due to IC scattering
within the base of a jet \citep[e.g.;][]{mat92,mar05}, is still under debate.
Observationally, the slope of the power-law continuum and its cutoff at high
energies provide direct information on the temperature and optical depth of the
corona, and, somewhat more loose constraints on its geometry
\citep{fab15,fab17}; and have an important effect on the shape of the
reflection spectrum \citep{gar13a,gar15b}.

\j1752\ is an X-ray transient discovered in 2009 October 23 by the All Sky
Monitor (ASM) on board the {\it Rossi X-ray Timing Explorer (RXTE)}
\citep{mar09}. Intense monitoring of its 2009 outburst (the only one detected
to date) with \rxte, \swift, and \maxi\ indicated that the source is a BHB
candidate \citep{mar09b,nak09,rem09,sha09,sha10,sha10b}. Radio and X-ray jets
have been detected in this source, including ballistic jets observed in the
radio, which are typically associated with hard-to-soft state transitions
\citep[generally during the intermediate or steep power-law
states;][]{bro10,yan10,yan11,bro13}. \cite{sha10} applied correlations between
spectral and variability properties for XTE~J1550--564 and GRO~J1655--40  in
order to  estimate a mass for \j1752\ of $9.6\pm0.8$~\msun, and a distance of
$3.5\pm0.4$~kpc. However, these quantities have not been verified and notably
there is no dynamical mass constraint. An inclination upper limit of $i<49$~deg
was found using photometric observations of the radio jet during the transition
from the hard to the soft state \citep{mil11}.

In this paper, we present a detailed spectroscopic analysis of the hard state
of \j1752.  \rxte\ pointed observations showed a protracted month-long period
in which the luminosity and hardness ratio of the source were found to be
extraordinarily stable, resulting in a unique dataset of exceptional quality
among stellar-mass black hole systems. Following the methodology developed
previously for \gx339\ \citep{gar15}, we combined these 300~ks of stable
hard-state data into a single spectrum with $\sim 100$~million source counts
($\sim$3--140~keV). Using a newly improved version of our relativistic
reflection model that includes a physical Comptonization continuum
\citep[\relxillCp;][]{gar18}, we derived precise constraints on the black hole
spin, the inner radius of the disk, the inclination of the reflector, the
ionization state and iron abundance of the disk's atmosphere, and the
temperature and optical depth of the corona.

This paper is organized as follows: In Section~\ref{sec:obs} we describe the
observational data; the spectral analysis is presented in
Section~\ref{sec:anal}.  We discuss the main implications of our results in
Section~\ref{sec:disc}, and offer concluding remarks in
Section~\ref{sec:concl}.
%

%
\graphicspath{{../plots/}}
\begin{figure}[ht!]
\centering
\includegraphics[width=\linewidth,trim={0 0 0 1cm}]{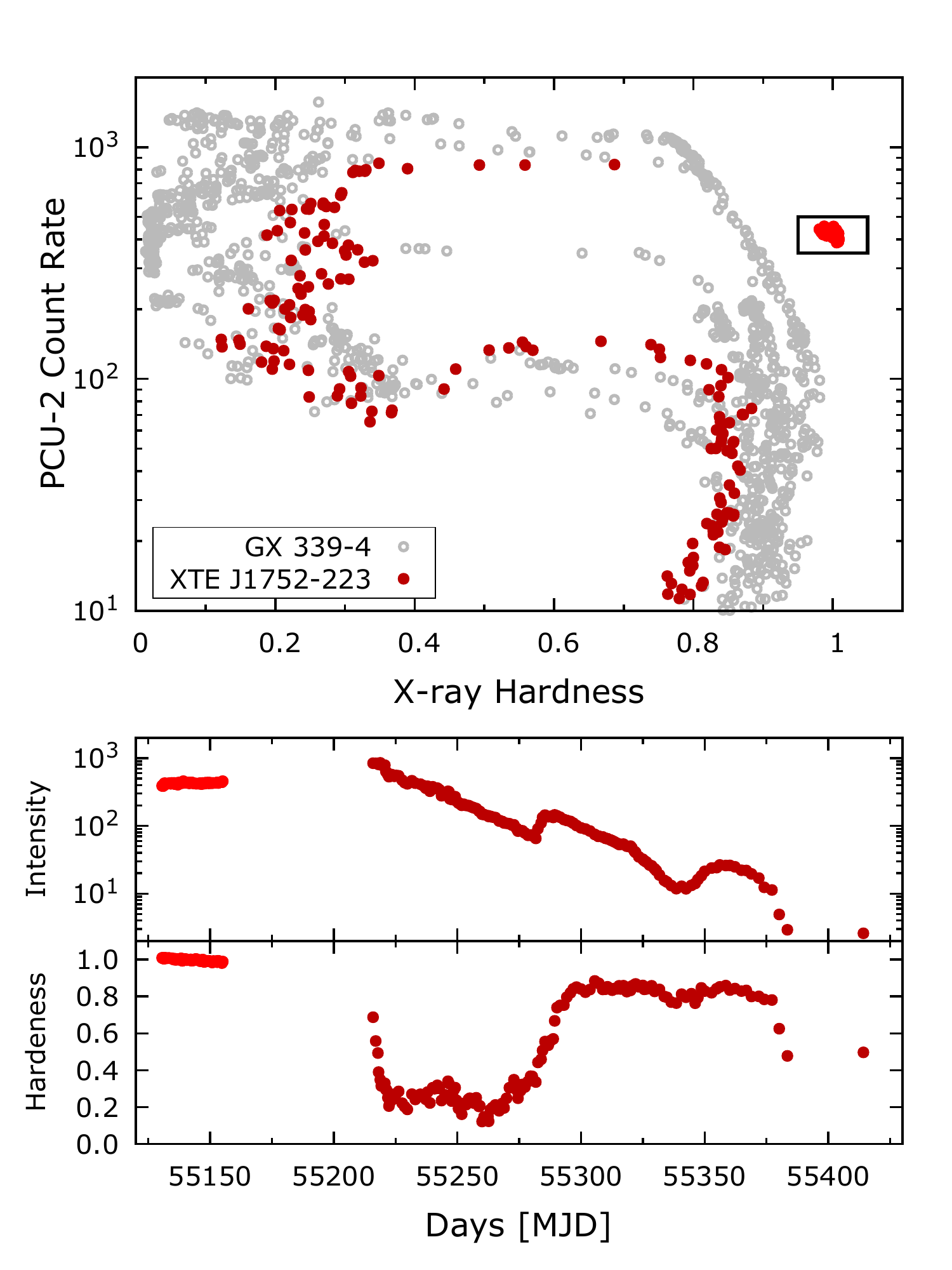}
\caption{({\it Top}) Hardness-intensity diagram for \gx339 (grey circles) and
\j1752 (red dots). The X-ray hardness in the horizontal axis is defined as the
ratio of the source counts at 8.6--18~keV to the counts at 5--8.6~keV
\citep{rem06}, after correcting to stabilize variations in the Crab rates to
account for gain drift. The PCU-2 count rate has been normalized to the Crab
following \cite{per16}, taking into account possible flux variations
\citep{wil11}.  All the hard-state data shown inside the box (light red) are combined into
a single, high signal-to-noise spectrum, equivalent to a $\sim 3\times 10^5$~s
exposure with a total of 100~million counts. ({\it Middle, Bottom}) PCA light
curve and hardness ratio as a function of time throughout the outburst. The
gap between MJDs $\sim 55160-55210$ shows a Sun exclusion period.
}
\label{fig:qdiagram}
\end{figure}

\section{Observations}\label{sec:obs}

We have analyzed the \rxte\ archival data for the hard state of \j1752,
specifically, all 6 observations from proposal ID~P94044, and the first 51
observations from proposal ID~P94331 starting with ObsID~94331-01-01-00 up to
ObsID~94331-01-05-00.  This includes spectra taken with the Proportional
Counter Array \citep[PCA;][]{jah06}, a set of five proportional counter
detectors sensitive over the energy range 2--60~keV; and spectra taken with the
the High Energy X-ray Timing Experiment \citep[HEXTE;][]{rot98}, a set of two
independent clusters (A and B), each with four NaI(Tl)/CsI(Na) phoswich
scintillation detectors sensitive over the 15--250~keV energy range.  We
followed the standard extraction procedure for both PCA and HEXTE as outlined
in \cite{Grinberg_2013a}, but used HEASOFT~6.16 and all Xenon layers for the
PCA extraction. We extracted one spectrum for each ObsID: for PCA, we used
\texttt{standard2f} PCU~2 spectra, discarding data within 10\,min from the
South Atlantic Anomaly. For HEXTE, we only extracted data from cluster~B due to
the failure of cluster~A earlier in the mission. As done in \cite{gar16}, the
HEXTE data was grouped beyond the standard reduction procedure by factors of 2,
3, and 4, in the energy ranges of 20--30~keV, 30--40~keV, and 40--250~keV,
respectively, in order to achieve an over-sampling of $\sim3$ times the
instrumental resolution.
%

%
\graphicspath{{../plots/}}
\begin{figure}[ht!]
\centering
\includegraphics[width=\linewidth]{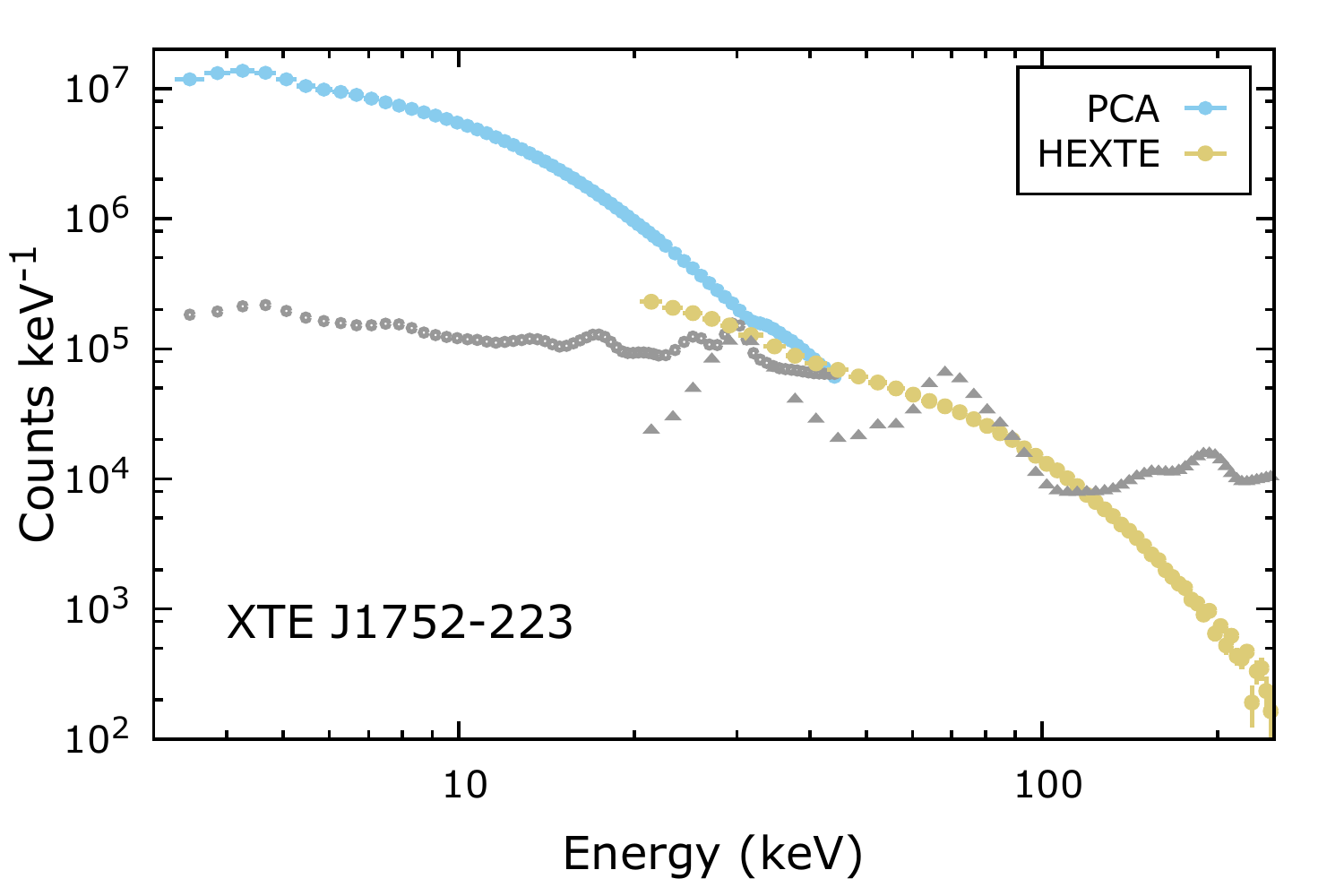}
\caption{Count spectra of the combined PCA (light blue) and HEXTE (gold) data for the
 hard-state of \j1752. The grey symbols show the background for each
 data set.
}
\label{fig:1752-lcounts}
\end{figure}

Figure~\ref{fig:qdiagram} (top) shows the standard hardness-intensity diagram
\citep{hom05} for the 2009 outburst of \j1752 (red points), together with the
multi-outburst data of the prototypical BHB \gx339 (grey circles). Unlike
\gx339, \j1752 displayed a nearly constant intensity (about 30\% of its peak
rate), for roughly a full month during the rise phase in the hard state,
corresponding to the concentration of points in the upper-right region of the
diagram, indicated inside the box (light red dots). These very stable
hard-state data were combined into a single spectrum following the procedures
described in \cite{gar15} for \gx339. A total of 57 individual pointings taken
during MJD 55130--55155 were combined into a unique dataset of exceptional
quality: a total of 300~ks were combined into a single PCA (3--45~keV) spectrum
with 100~million counts, and a single HEXTE spectrum (20--250~keV) with
10~million counts. Middle and lower panels of Figure~\ref{fig:qdiagram} show
the light curve and hardness ratio as function of time for \j1752, respectively.
The gap between the hard-state observations (light red) and the transition to
the soft state (darker red), is due to a Sun exclusion period. 
%

%
\graphicspath{{../plots/}}
\begin{figure*}[ht!]
\centering
\includegraphics[width=0.8\linewidth,trim={0 20pt 0 0}]{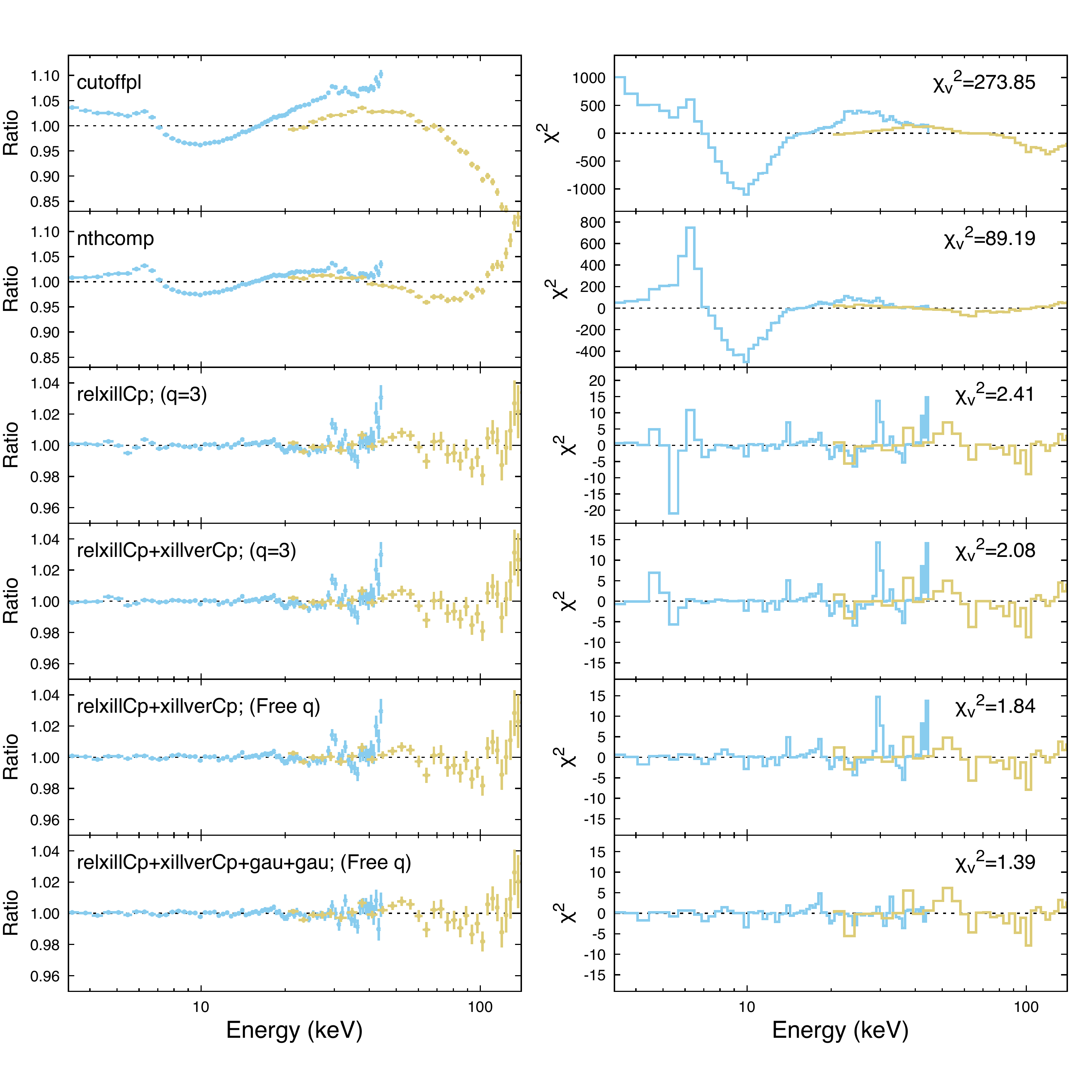}
\caption{Ratio plot (left) and contribution to $\chi^2$ (right) from the
different model fits. The model used is show in the left panels, and the
corresponding $\chi_{\nu}^2$ is indicated in the right panels.
}
\label{fig:progression}
\end{figure*}

The final PCA spectrum was further calibrated with our correction tool
\pcacorr\ \citep{gar14b}, which improves the data quality and accordingly
enhances the detector sensitivity to more subtle spectral features such as the Fe
K line and edge. Mere 0.1\% systematic uncertainties are sufficient after this
correction for analyzing the PCA dataset. Given that all the HEXTE observations
for \j1752 were taken with the Cluster~B, we have also corrected the final
HEXTE spectrum with the \hexBcorr\ tool, as described in \cite{gar16}. No
systematics are included to the HEXTE spectra.

The net spectra for both PCA and HEXTE are shown in
Figure~\ref{fig:1752-lcounts}, including their corresponding backgrounds.  The
HEXTE background becomes dominant at high energies, above $\sim$100~keV.  At
250~keV the background is more than a factor of ten higher than the source
counts. We thus limit our analysis up to 140~keV, where the background counts
are no higher than 50\% the source counts. In the analyzed HEXTE range
(20--140~keV), there are 5.8~million counts.

\vspace{40pt}

\section{Spectral Analysis}\label{sec:anal}

\subsection{Exploration: Empirical Determination of the Model Components}\label{sec:deter}

The spectral analysis of the combined hard-state data for \j1752\ is carried
out by simultaneously fitting the PCA and HEXTE spectra.  For the PCA spectra,
channels 1--4 and energies above 45~keV are ignored. For the HEXTE spectra, we
only consider the 20--140~keV range. The fitting and statistical analysis was carried out using
the {\sc xspec} package v-12.9.0d \citep{arn96}.

The present analysis follows closely our previous work on the hard state of
\gx339 \citep{gar15}. However, there are two important methodological
differences here: (i) we included simultaneous high-energy data provided by
HEXTE, which has been corrected with our \hexBcorr\ tool \citep{gar16}; and
(ii) we have updated our reflection models to now include a physically
motivated Comptonization continuum.
%

%
\begin{table*}
\begin{center} 
\small
\caption{Statistics for the progression of initial model fits}
\begin{tabular}{lrcrr}
\hline
Model & $\chi^2$ & $\nu$ & $\chi^2_{\nu}$ & $\Delta\chi^2/\Delta\nu$ \\
\hline
{\tt const*TBabs*cutoffpl}                                  & 27932.32 & 102 & 273.85 &          \\
{\tt const*TBabs*nthComp}                                   &  9097.31 & 102 &  89.19 & 18835.01 \\
{\tt const*TBabs*relxillCp}                                 &   233.83 &  97 &   2.41 &  8863.48 \\
{\tt const*TBabs*(relxillCp+xillverCp)}; ($q=3$)            &   199.32 &  96 &   2.08 &    34.51 \\
{\tt const*TBabs*(relxillCp+xillverCp)}; (Free $q$)         &   175.00 &  95 &   1.84 &    24.32 \\
{\tt const*TBabs*(relxillCp+xillverCp+gau+gau)}; (Free $q$) &   127.49 &  92 &   1.39 &    15.70 \\
\hline
\end{tabular}
\label{tab:progression}
\end{center}
\end{table*}
%

Figure~\ref{fig:progression} shows the residuals resulting from a progression
of models applied to the \j1752\ data, which sequentially increment in
complexity.  The statistics of each fit are summarized in
Table~\ref{tab:progression}. For all models, a normalization constant is
included to account for the differences in the flux calibration between the PCA
and the HEXTE instruments. Galactic absorption is included by implementing the
{\tt TBabs} model with the corresponding abundances as set by \cite{wil00}, and
the \cite{ver96} photoelectric cross sections.

We first start with a simple model for the continuum in the form of a power law
with an exponential cutoff at high energies (i.e., {\tt cutoffpl}).  Very large
residuals can be seen in the top panels of Figure~\ref{fig:progression}, which
resemble the signatures of reprocessing from an optically-thick material in the
form of a broad Fe K emission line at $\sim$6.6~keV, a smeared Fe K edge at
$\sim$8~keV, and a broad Compton hump peaking at $\sim$30~keV. Despite the
inclusion of a cross calibration constant, this model fails to correctly
describe the curvature at high energies and thus there appears to be a mismatch
between the two spectra.

The presence of a power-law continuum with a high-energy cutoff is commonly
attributed to the emission from an optically-thin, hot Comptonizing corona
\citep[e.g.;][]{don07}. Thus, we replace the e-folded power-law model with a
physically motivated thermal Comptonization model \nthComp
\footnote{\url{https://heasarc.gsfc.nasa.gov/xanadu/xspec/models/nthcomp.html}},
included as part of {\sc xspec}. This model, developed by \cite{zdz96} and
later extended by \cite{,zyc99}, provides a more accurate description of the
cutoff at high energies, which is sharper than the exponential cutoff.
In this prescription, the seed photons from
the thermal disk emission (a quasi blackbody) are Compton up-scattered by the
hot electrons in the corona.  The residuals of this fit are shown in the second
panels of Figure~\ref{fig:progression}.  This model provides a significantly
better match to the data bringing the reduced chi-square from 274 to 89, and
providing a better agreement between the two datasets in the spectral region
where they overlap (20--45~keV). In this case, the high energy cutoff is much
sharper than the exponential, which not only affects the shape of the continuum
but also the shape of the reflected spectrum, as we describe next.

To model the residuals observed we then make use of our suite of relativistic
reflection models \relxill\ ({\tt v-0.4j})\footnote{\url{
http://www.sternwarte.uni-erlangen.de/research/relxill}}
\citep{gar14a,dau14}.  This model is the result of the merging of the ionized
reflection spectra produced with the {\sc xillver} code
\citep{gar10,gar11,gar13a}, with the ray tracing calculations based on the
relativistic convolution kernel {\sc relline} \citep{dau10,dau13}.  The \relxill\
models properly take into account the angular dependence of the reflection as a
function of the radius in the accretion disk, including the most recent dataset
of atomic quantities. The relativistic effects that smear and modify the
spectrum are included considering two basic geometries: the {\it extended corona}
(in the standard \relxill), assuming that the emissivity of the disk follows a
powerlaw with the radius $\propto r^{-q}$, with the index $q$ being a fit parameter;
and the {\it lamppost corona} (in \relxilllp), assuming a point-like source at
the rotation axis above the black hole (with the height $h$ being a fit
parameter).  In all cases, the model provides both the illuminating continuum
and the reflected spectrum for a given set of parameters. The shape of the
continuum can be a powerlaw with an exponential cutoff at high energies (which
is the default in all the model flavors), or a thermal Comptonization continuum
(in all the flavors with the {\tt Cp} nomenclature), as described below.

Given the dramatic improvement in the fit achieved by the use of the \nthComp\
continuum, we have implemented the new version of our relativistic reflection
model \relxillCp, in which the reflection spectrum is self-consistently
calculated using the more physical illumination continuum calculated with \nthComp.
This model has the same number of parameters as the earlier version
\relxill\ (which uses an e-folded power-law continuum), with the only
difference that the high-energy cutoff parameter $E_\mathrm{cut}$ is now
replaced by the coronal electron temperature $kT_e$. A typical correspondence
between the cutoff prescriptions is $E_\mathrm{cut}\sim (2-3)kT_e$, depending
on the optical depth and geometry of the corona. The addition of this component
results in a dramatic improvement of the fit, with the reduced chi-square
changing from 89 to 2.4. For this fit the emissivity profile is assumed to
follow a power law with an index fixed at the canonical value of $q=3$.

While most of the reflection features are well modeled by \relxillCp, some
residuals still remain in the Fe K region near 6-7~keV. These residuals are
plausibly due to an unmodelled narrow line component.  Thus, we have also
included an unblurred reflection component to our fits (fourth panels in
Figure~\ref{fig:progression}), similar to our previous fits to \gx339.
However, once again we implement our new reflection models produced with a
thermal Comptonization continuum, i.e., the \xillverCp\ model. All parameters
are linked between the \relxillCp\ and the \xillverCp\ components, with
exception of the ionization parameter which is fixed at its lowest value
($\log\xi=0$), assuming that the material is nearly neutral; and the reflection
fraction, which is let free to vary but constrained to negative values (a
setting option so that no continuum component is added in).  For \gx339, the
data were strongly incompatible with a linked Fe abundance between the narrow
and broad reflection components, but here we find that there is no empirical
need for decoupling those abundances.  Therefore, only one additional free
parameter is introduced by including \xillverCp. The residuals near the Fe K
region are significantly minimized ($\Delta\chi^2_{\nu}=34.5$), although not
completely removed. An even better fit is found using the same model but
allowing for the emissivity index $q$ to be free. The improvement in the fit
statistics is significant ($\Delta\chi^2_{\nu}=24.3$), and all the residuals in
the Fe K region are minimized (second to last panels in
Figure~\ref{fig:progression}). 

Two relatively large residuals are still observed at $\sim$30~keV and
$\sim$42~keV, which are only present in the PCA spectra but absent in the
HEXTE. This suggests that origin of these features could be instrumental. It is
possible that \pcacorr\ does not fully reduce instrumental features at these
energies since it is based on the analysis of Crab data, which has a much
softer spectrum ($\Gamma\sim 2.1$) than \j1752\ ($\Gamma\sim 1.5$). Therefore,
two {\it ad hoc} Gaussian profiles are included in our model (but only effective
to the PCA data), with their widths
fixed at $\sigma=0.1$~keV. The energy of the first Gaussian is fixed at
29.8~keV, which corresponds to one of the $^{241}$Am radioactive emission lines
\citep{jah06}. The energy of the second Gaussian is constrained to the
40--45~keV range and fitted for. The residuals of this model are shown in
the last panel of Figure~\ref{fig:progression}. The inclusion of these two
Gaussians has no effect on the other model parameters, and their effect is
merely cosmetic (i.e., to improve the fit quality).

\subsection{Spectral Fits}\label{sec:fits}

The progression of different model components described in the previous section
demonstrates that a model composed of a thermal Comptonization continuum,
relativistic and non-relativistic reflection (in addition to the two cosmetic
Gaussians), provides a very good description of our hard-state data for \j1752.
With the above exploratory analysis guiding our approach, we next apply three
different model fits aimed to determine the physical properties of this system.

We first start by replacing the simple cross correlation constant with a
natural extension that allows for both normalization and shape differences via
the model \crabcorr\ \citep{ste10}. This model is designed to standardize
detector responses to return the same normalizations and power-law slopes for
the Crab. We adopt as our standard, the \cite{too74} spectral fit (i.e.,
$\Gamma=2.1$ and $N=9.7$~photons~s$^{-1}$~keV$^{-1}$ at 1~keV). {\tt Crabcorr}
multiplies a model spectrum by a power law, applying both normalization ($N$)
and “tilt” ($\Delta\Gamma$) corrections.  These quantities are frozen at the
measured values for the Crab based on PCA data \citep[i.e.; $N=1.097$ and
$\Delta\Gamma=0.01$;][Table 1]{ste10}, and left free to vary for the HEXTE
data.  
%

\def\mInh{$1.00\pm 0.11$}                
\def\mIq{$>7.2$}                         
\def\mIi{$67^{+2}_{-8}$}                 
\def\mIrin{$1.13^{+0.13}_{-0.06}$}       
\def\mIga{$1.548^{+0.009}_{-0.006}$}     
\def\mIxi{$3.05^{+0.12}_{-0.07}$}        
\def\mIafe{$3.6 \pm 0.4$}                
\def\mIkT{$59 \pm 3$}                    
\def\mIrrf{$0.19 \pm 0.03$}              
\def\mIrn{$(2.7\pm0.03)\times10^{-9}$}   
\def\mIrf{$(2.9\pm0.5)\times10^{-3}$}    
\def\mIgeii{$43.2 \pm 0.9$}              
\def\mIgni{$2.1 \pm 0.8$}                
\def\mIgnii{$1.5 \pm 0.6$}               
\def\mIdg{$3 \pm 8$}                     
\def\mIdn{$0.90 \pm 0.02$}               
\def\mIchi{$137$}
\def\mIdof{$90$}
\def\mIredc{$1.522$}

\def\mIIAnh{$1.12 \pm 0.09$}                 
\def\mIIAh{$2.0^{+0.8}_{-0.3}$}              
\def\mIIAi{$35^{+3}_{-5}$}                   
\def\mIIArin{$1.4^{+0.4}_{-0.2}$}            
\def\mIIAga{$1.545^{+0.003}_{-0.005}$}       
\def\mIIAxi{$3.11 \pm 0.07$}                 
\def\mIIAafe{$3.7^{+0.6}_{-0.4}$}            
\def\mIIAkT{$57.1 \pm 1.9$}                  
\def\mIIArrf{$0.30^{+0.10}_{-0.07}$}         
\def\mIIArn{$(1.6 \pm 0.5) \times 10^{-9}$}  
\def\mIIArf{$-0.28^{+ 0.06}_{- 0.08} $}      
\def\mIIAgeii{$43.2 \pm 0.9$}                
\def\mIIAgni{$2.0 \pm 0.8$}                  
\def\mIIAgnii{$1.5^{+0.7}_{-0.5}$}           
\def\mIIAdg{$6 \pm 6$}                       
\def\mIIAdn{$0.906 \pm 0.019 $}              
\def\mIIAchi{$143$}                             
\def\mIIAdof{$90$}                              
\def\mIIAredc{$1.584$}                          

\def\mIIBnh{$1.10 \pm 0.08$}                  
\def\mIIBh{$1.15^{+0.38}_{-0.06}$}            
\def\mIIBa{$0.92^{+0.05}_{-0.07}$}            
\def\mIIBi{$36 \pm 4$}                        
\def\mIIBga{$1.546 \pm 0.004$}                
\def\mIIBxi{$3.09^{+0.10}_{-0.05}$}           
\def\mIIBafe{$3.6 \pm 0.5$}                   
\def\mIIBkT{$57.4 \pm 1.9$}                   
\def\mIIBrrf{$0.20^{+0.07}_{-0.11}$}          
\def\mIIBrn{$(1.6 \pm 0.6) \times 10^{-9}$}   
\def\mIIBrf{$-0.31 \pm 0.07$}                 
\def\mIIBgeii{$43.2^{+1.3}_{-0.7}$}           
\def\mIIBgni{$2.0 \pm 0.9$}                   
\def\mIIBgnii{$1.5 \pm 0.6$}                  
\def\mIIBdg{$5 \pm 7$}                        
\def\mIIBdn{$0.903 \pm 0.021$}                
\def\mIIBchi{$142$}                           
\def\mIIBdof{$90$}
\def\mIIBredc{$1.577$}

\def\mIIIAnh{ $1.0 \pm 0.1 $}      
\def\mIIIAh{ $ 1.17^{+ 0.85}_{- 0.07} $}      
\def\mIIIArin{ $1.8 \pm 0.4 $}      
\def\mIIIAi{ $ 31 \pm 6 $}      
\def\mIIIAga{ $ 1.62^{+ 0.02}_{- 0.03} $}      
\def\mIIIAxi{ $ 3.24^{+ 0.09}_{- 0.13} $}      
\def\mIIIAafe{ $3.3^{+ 0.7}_{- 0.4} $}      
\def\mIIIAkT{ $ 70 \pm 6 $}      
\def\mIIIArrf{ $ 3.5^{+ 12.4}_{- 2.8} $}      
\def\mIIIArn{ $ 3^{+ 2}_{- 1} \times 10^{-9} $}      
\def\mIIIArf{ $ (5 \pm 2 ) \times 10^{5} $}      
\def\mIIIAgeii{ $ 43.5^{+ 1.2}_{- 0.6} $}      
\def\mIIIAgni{ $2.0 \pm 0.9 $}      
\def\mIIIAgnii{ $1.6 \pm 0.5 $}      
\def\mIIIAdg{ $5^{+5}_{- 7} $}      
\def\mIIIAdn{ $0.91^{+ 0.01}_{- 0.03} $}      
\def\mIIIAfsc{ $ 0.83^{+ 0.02}_{- 0.08} $}      
\def\mIIIAchi{117}               
\def\mIIIAdof{$89$}
\def\mIIIAredc{$1.315$}

\def\mIIIBnh{ $1.00 \pm 0.13 $}      
\def\mIIIBh{ $1.4^{+ 0.4}_{- 0.3} $}      
\def\mIIIBa{ $ 0.95^{+ 0.04}_{- 0.13} $}      
\def\mIIIBi{ $30^{+ 6}_{- 8} $}      
\def\mIIIBga{ $ 1.62^{+ 0.02}_{- 0.03} $}      
\def\mIIIBxi{ $ 3.17^{+ 0.15}_{- 0.07} $}      
\def\mIIIBafe{ $3.4^{+ 0.8}_{- 0.5} $}      
\def\mIIIBkT{ $ 70 \pm 6 $}      
\def\mIIIBrrf{ $ 6.9^{+ 23.1}_{- 5.8} $}      
\def\mIIIBrn{ $ 3^{+ 2}_{- 1} \times 10^{-9} $}      
\def\mIIIBrf{ $ 4^{+ 3}_{- 2} \times 10^{5} $}      
\def\mIIIBgeii{ $ 43.4^{+ 1.3}_{- 0.5} $}      
\def\mIIIBgni{ $2.0 \pm 0.9 $}      
\def\mIIIBgnii{ $1.5 \pm 0.6 $}      
\def\mIIIBdg{ $6^{+ 4}_{- 8} $}      
\def\mIIIBdn{ $0.90 \pm 0.02 $}      
\def\mIIIBfsc{ $ 0.82^{+ 0.03}_{- 0.07} $}      
\def\mIIIBchi{116}               
\def\mIIIBdof{$89$}
\def\mIIIBredc{$1.303$}

%
\begin{table*}
\begin{center}
\small
\caption{Best-fit parameters for the final fits with relativistic reflection modeling.}
\begin{tabular}{lcrrrrrrr}
\hline
Component & Parameter & Model~1 & Model~2.A & Model~2.B & Model~3.A & Model~3.B\\
\hline
{\tt TBabs}         & $N_\mathrm{H}$ ($10^{22}$ cm$^{-2}$)& \mInh\   & \mIIAnh\   & \mIIBnh\   & \mIIIAnh\   & \mIIIBnh\   \\
{\tt relxill(lp)Cp} & $a_*$                               & $0.998$  & $0.998$    & \mIIBa\    & $0.998$     & \mIIIBa\    \\
{\tt relxillCp}     & $q$                                 & \mIq\    & \nodata    & \nodata    & \nodata     & \nodata     \\
{\tt relxilllpCp}   & $h$ $(R_\mathrm{Hor})$              & \nodata  & \mIIAh\    & \mIIBh\    & \mIIIAh\    & \mIIIBh\    \\
{\tt relxill(lp)Cp} & $i$ (deg)                           & \mIi\    & \mIIAi\    & \mIIBi\    & \mIIIAi\    & \mIIIBi\    \\
{\tt relxill(lp)Cp} & $R_\mathrm{in}$ ($R_\mathrm{ISCO})$ & \mIrin\  & \mIIArin\  & $1$        & \mIIIArin\  & $1$         \\
{\tt relxill(lp)Cp} & $\Gamma$                            & \mIga\   & \mIIAga\   & \mIIBga\   & \mIIIAga\   & \mIIIBga\   \\
{\tt relxill(lp)Cp} & $\log\xi$ (erg cm s$^{-1}$)         & \mIxi\   & \mIIAxi\   & \mIIBxi\   & \mIIIAxi\   & \mIIIBxi\   \\
{\tt relxill(lp)Cp} & $A_\mathrm{Fe}$                     & \mIafe\  & \mIIAafe\  & \mIIBafe\  & \mIIIAafe\  & \mIIIBafe\  \\
{\tt relxill(lp)Cp} & $kT_e$ (keV)                        & \mIkT\   & \mIIAkT\   & \mIIBkT\   & \mIIIAkT\   & \mIIIBkT\   \\
{\tt relxill(lp)Cp} & $R_\mathrm{f}$                      & \mIrrf\  & \mIIArrf\  & \mIIBrrf\  & \mIIIArrf\  & \mIIIBrrf\  \\
{\tt relxill(lp)Cp} & $N$                                 & \mIrn\   & \mIIArn\   & \mIIBrn\   & \mIIIArn\   & \mIIIBrn\   \\
{\tt xillverCp}     & $N$                                 & \mIrf\   & \mIIArf\   & \mIIBrf\   & \mIIIArf\   & \mIIIBrf\   \\
{\tt Gaussian 1}    & $E$ (keV)                           & $29.8$   & $29.8$     & $29.8$     & $29.8$      & $29.8$      \\
{\tt Gaussian 1}    & $N$ $(10^{-4})$                     & \mIgni\  & \mIIAgni\  & \mIIBgni\  & \mIIIAgni\  & \mIIIBgni\  \\
{\tt Gaussian 2}    & $E$ (keV)                           & \mIgeii\ & \mIIAgeii\ & \mIIBgeii\ & \mIIIAgeii\ & \mIIIBgeii\ \\
{\tt Gaussian 2}    & $N$ $(10^{-4})$                     & \mIgnii\ & \mIIAgnii\ & \mIIBgnii\ & \mIIIAgnii\ & \mIIIBgnii\ \\
{\tt crabcorr}      & $\Delta\Gamma$ $(10^{-3})$          & \mIdg\   & \mIIAdg\   & \mIIBdg\   & \mIIIAdg\   & \mIIIBdg\   \\
{\tt crabcorr}      & $N$                                 & \mIdn\   & \mIIAdn\   & \mIIBdn\   & \mIIIAdn\   & \mIIIBdn\   \\
{\tt simplcut}      & $f_{sc}$			          & \nodata  & \nodata    & \nodata    & \mIIIAfsc\  & \mIIIBfsc\  \\
\hline
$\chi^2$            & \nodata                             & \mIchi\  & \mIIAchi\  & \mIIBchi\  & \mIIIAchi\  & \mIIIBchi\  \\
$\nu$               & \nodata                             & \mIdof\  & \mIIAdof\  & \mIIBdof\  & \mIIIAdof\  & \mIIIBdof\  \\
$\chi_{\nu}^2$      & \nodata                             & \mIredc\ & \mIIAredc\ & \mIIBredc\ & \mIIIAredc\ & \mIIIBredc\ \\
\hline
\end{tabular}
\label{tab:fits}
\end{center}
\end{table*}

The first model is essentially the same final model described in
Section~\ref{sec:deter}, with the replacement of the cross calibration constant
by the \crabcorr\ model. The subsequent 4 models assume a lamppost geometry
(i.e., a point source corona on the spin axis at a height $h$ above the disk)
for the relativistic reflection, which is achieved by replacing \relxillCp\ by
\relxilllpCp. These models are divided into two classes (2 \& 3) and two
sub-cases (A \& B). Models~2.A and 2.B adopt \relxilllpCp\ plus the unblurred
reflection component (\xillverCp). Models~3.A and 3.B include the model
\simplcut, which accounts for the Comptonization of reflected emission in the
corona \citep[see][for a detailed discussion of the model]{ste17}. In case A,
we fit for the inner radius while keeping the spin fixed at the Thorne limit
\citep[$a_*=0.998$;][]{tho74}, in cases B we assume the inner radius
corresponds to the inner-most stable circular orbit (ISCO), and fit for the spin.
All five models are then written as:
\begin{itemize}
\item {\bf Model~1 (fixed spin):} \\
{\tt crabcorr*TBabs*(relxillCp+xillverCp+gau+gau)};

\item {\bf Model~2.A (fixed spin):} \\
{\tt crabcorr*TBabs*(relxilllpCp+xillverCp+gau+gau)};

\item {\bf Model~2.B (fixed $R_\mathrm{in}$):} \\
{\tt crabcorr*TBabs*(relxilllpCp+xillverCp+gau+gau)};

\item {\bf Model~3.A (fixed spin):} \\
{\tt crabcorr*TBabs*(simplcut*relxilllpCp + xillverCp + gau + gau + nthComp)};

\item {\bf Model~3.B (fixed $R_\mathrm{in}$):}\\
{\tt crabcorr*TBabs*(simplcut*relxilllpCp + xillverCp + gau + gau + nthComp)};
\end{itemize}
%

%
\graphicspath{{../plots/}}
\begin{figure*}[ht!]
\centering
\includegraphics[width=0.7\linewidth,trim={0 0 0 0}]{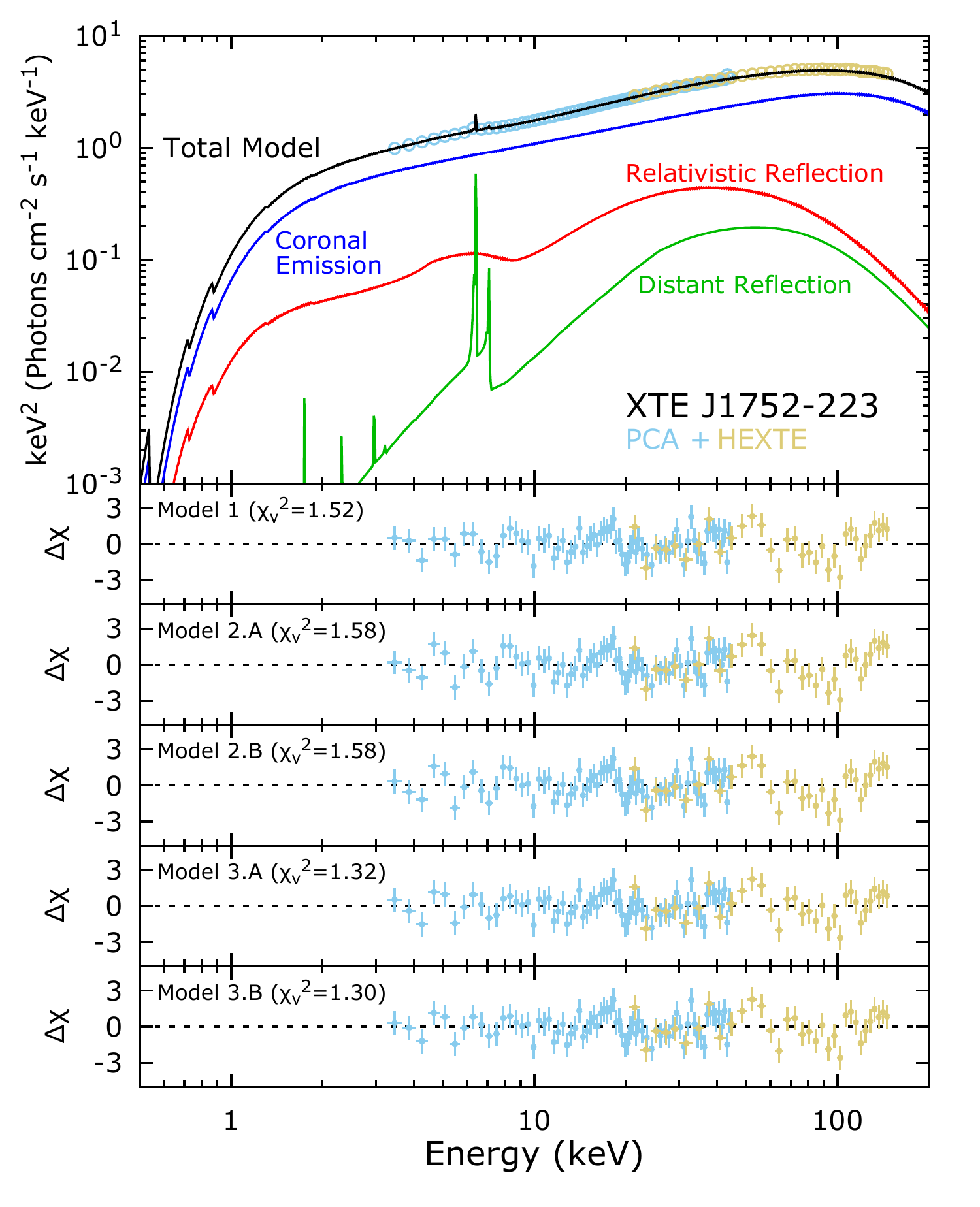}
\caption{Model components and residuals for the  advanced fits using reflection
modeling.  {\it Top:} Unfolded spectra for the PCA (light blue) and HEXTE
(gold) combined hard-state data of \j1752. For the best fit with Model~1, the
components shown are: coronal emission, modeled with {\tt nthcomp} (blue);
relativistic reflection, modeled with \relxillCp\ (red); and distant
(unblurred) reflection, modeled with \xillverCp\ (green). The total model is
shown in black, and all these components include Galactic absorption modeled
with {\tt TBabs}. {\it Bottom panels:} residuals from the best fits using
Models~1, 2.A, 2.B, 3.A, and 3.B, indicating their corresponding
fit-statistics.
}
\label{fig:J1752-model-all}
\end{figure*}

The results from these five fits are summarized in Table~\ref{tab:fits}, and
the the model components obtained for Model~1 (which are very similar in the
other fits), together with the residuals of the five models are shown in
Figure~\ref{fig:J1752-model-all}.  The fit statistics are acceptable in all the
fits, in particular if one considers the remarkably large number of source
counts in these observations (about 100~million overall), and the very low
systematics (0.1\%).  Model~1 appears to be slightly better than Model~2 based
on the statistics, however, the improvement over the lamppost version is only
marginal ($\Delta\chi^2 = 5-6$, with respect to Models~2.B--2.A, respectively).
The inclusion of the extra Comptonization of the reflected component in
Models~3 results in a much more significant improvement, with a decrease of
$\Delta\chi^2 = 26$ with respect to Models~2, with the addition of only one
extra free parameter.  The differences, although statistically significant, are
difficult to discern by eye from the residuals shown in the lower panels of
Figure~\ref{fig:J1752-model-all}.  This is once again a consequence of the very
high signal-to-noise ratio of this dataset.
%

%
\graphicspath{{../plots/}}
\begin{figure*}[ht!]
\centering
\includegraphics[width=0.9\linewidth]{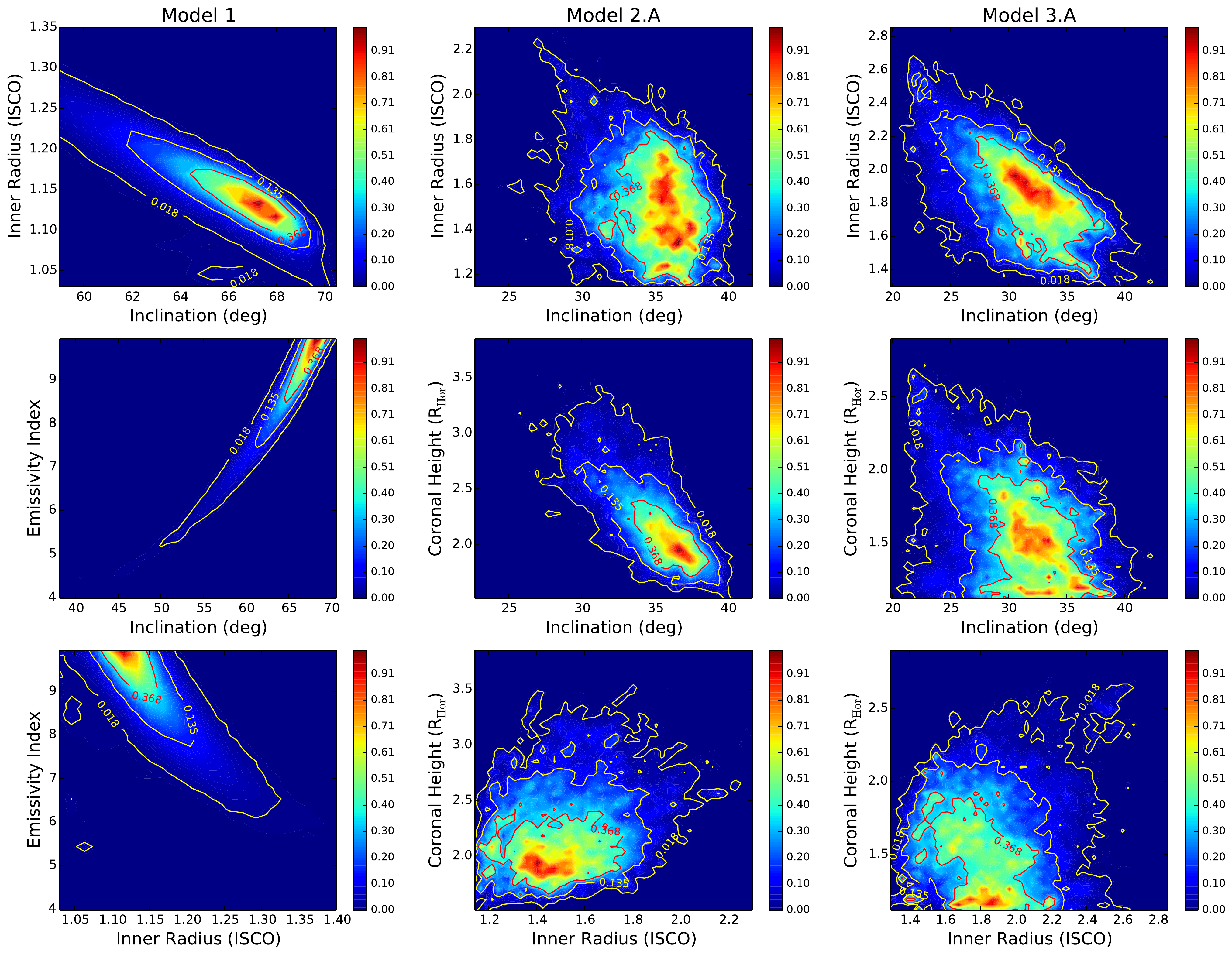}
\caption{Probability contours derived from the MCMC analysis of the fits using
free emissivity index (Model~1, left panels), and with the lamppost geometry
with free inner-radius (Model~2.A, right panels).  A selection of important
parameters is shown: inner radius, inclination, emissivity index, and coronal
height. The strongest degeneracies are seeing among parameters in Model~1.
}
\label{fig:plots-1}
\end{figure*}

Given the complexity of the reflection models adopted here, the statistical
analysis of all the fits, including the uncertainties of the parameters quoted
in Table~\ref{tab:fits}, was achieved by implementing a Markov Chain
Monte-Carlo (MCMC) algorithm. Specifically, we used the {\sc emcee-hammer}
Python package \citep{for13}, which allows efficient exploration of complex
parameter spaces in determining posterior probability distributions. Each MCMC
run consisted of 100--128 walkers (distinct chains), which was run until
convergence was reached. Convergence was assessed by requiring that for each
parameter at least 12 autocorrelation lengths were traversed by the average
walker. The first third of the run was then discarded as burn-in phase.
%

%
\graphicspath{{../plots/}}
\begin{figure*}[ht!]
\centering
\includegraphics[width=0.9\linewidth]{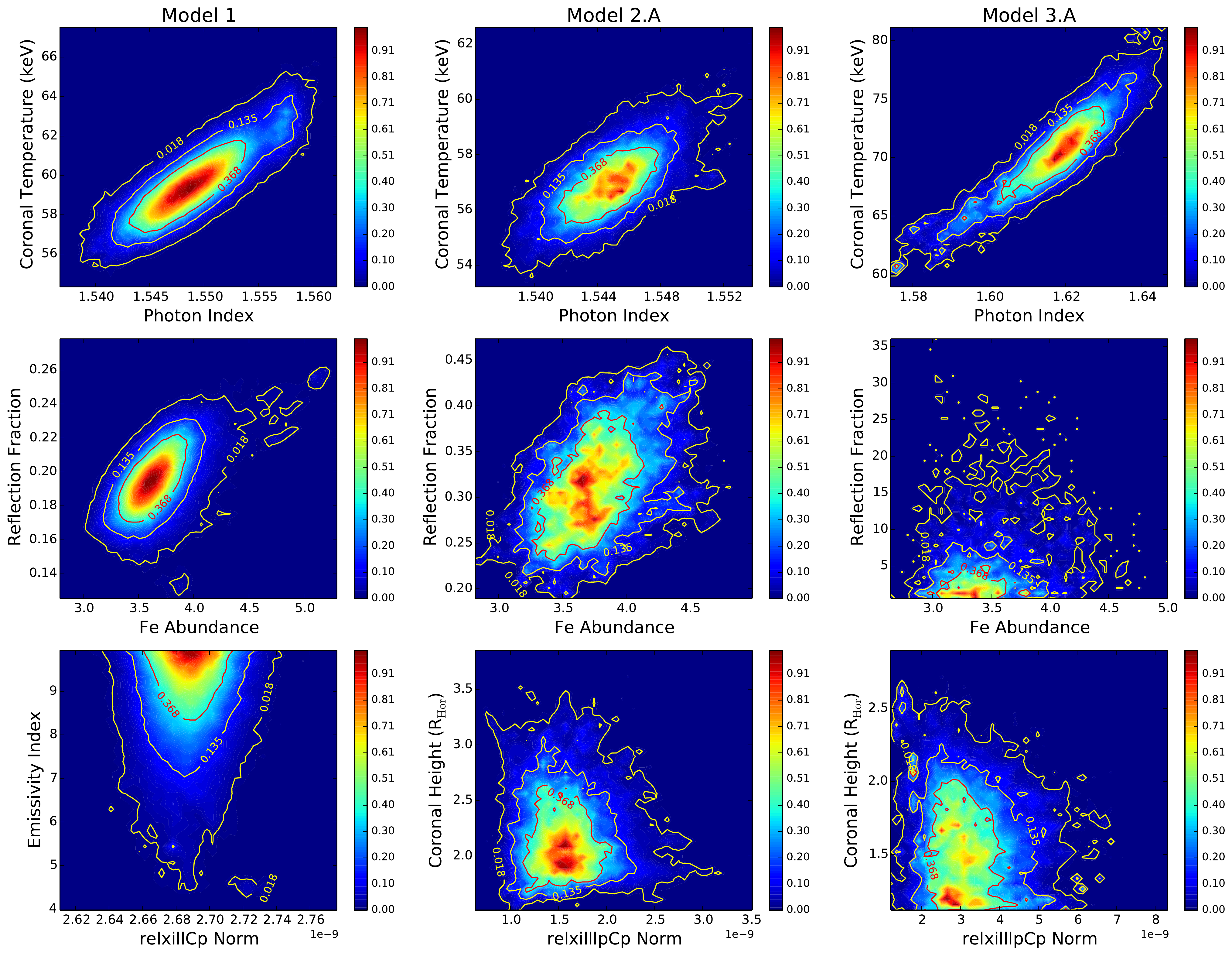}
\caption{Probability contours derived from the MCMC analysis of the fits using
free emissivity index (Model~1, left panels), and with the lamppost geometry
with free inner-radius (Model~2.A, right panels).  A selection of important
parameters is shown: photon index, coronal height, emissivity index, reflection
fraction, Fe abundance, and the normalization of the relativistic reflection
component. Most of these contours are consistent with each other, and no strong
degeneracies are observed.
}
\label{fig:plots-2}
\end{figure*}

\section{Discussion}\label{sec:disc}

In Figures~\ref{fig:plots-1} and \ref{fig:plots-2} we show the contour maps
derived via MCMC analysis for a set of the most relevant physical parameters.
For each map we also show the 1-, 2-, and 3-$\sigma$ confidence contours. These
maps illustrate how well these parameters are constrained in each fit, and the
level of degeneracy between parameters. In particular, we can see clear
correlations between the inner radius, inclination, and the emissivity index
(in the case of Model~1), or the height of the corona (in the case of
Model~2.A). These correlations appear to be much stronger for Model~1 than for
Model~2.A and Model~3.A (Figure~\ref{fig:plots-1}). We also observe the
expected correlation between the coronal temperature and the photon index in
both fits, while other important parameters such as reflection fraction and Fe
abundance, or the emissivity/coronal height and the normalization, show little
dependence on each other (Figure~\ref{fig:plots-2}).

Despite the fact that Model~1 yields a slightly better fit than Model~2.A, it
does not necessarily provide the best physical interpretation of the data. In
Model~1, the relativistic reflection component is assumed to follow an
emissivity profile in the form of a power law in the radial coordinate. In
other words, the net reflected emission $F_\mathrm{ref}$ follows
$F_\mathrm{ref} \propto r^{-q}$.  As noted in the progression of models of
Section~\ref{sec:deter}, the fit with Model~1 requires the emissivity index to
be very large, essentially pegging the parameter at its maximum value of 10. At
90\% confidence, a lower limit is found at \mIq.

The very steep emissivity found with Model~1 suggests an extreme relativistic
scenario where the illumination is compact and concentrated at the central
regions. This motivated the application of lamppost geometry in Models~2 and
3, where the \relxillCp\ component is replaced by \relxilllpCp.

In Models~2.A \& 3.A the spin is also fixed at the Thorne limit, while
$R_\mathrm{in}$ is let free to vary. Thus, these fits can be directly compared
to the results from Model~1. Between Model~2.A and Model~1, all parameters are
consistent within their uncertainties, with two exceptions: The reflection
fraction in Model~2.A is roughly 1.5 times the value found with Model~1. The
differences are more pronounced when comparing to Model~3.A. In the Model~3
variants, the inclusion of coronal Comptonization results in a reflection
fraction an order of magnitude larger than in Model~2 or Model~1. (In fact,
$R_f$ is distributed approximately log-normal, with a 99\% lower limit of
$>0.3$ for Model~3.A.)  Because Comptonization hardens the reflection output,
the intrinsic emission is found to be significantly softer $\Delta\Gamma
\approx 0.07$ and the electron temperature is likewise higher ($\Delta kT_e
\approx 12$ keV).

Most importantly, Models~2 and 3 strongly disagree with Model~1 on the
inclination of the system: the best-fit value for Model~1 is \mIi~deg, while
for Model~2.A and ~3.A the values are \mIIAi~deg and \mIIIAi~deg, respectively.
Meanwhile, \cite{mil11} found an upper limit of the inclination of $i < 49$~deg
based on radio observations of the jet when the source was transitioning from
the hard to the soft state. This upper limit thus formally excludes the large
inclination value from Model~1. It is possible that the extreme relativistic
effects are forcing the fit in Model~1 to increase the emissivity index to
unphysical values.\footnote{For instance, this value exceeds the maximum
emissivity predicted by a lamppost geometry for the given value of $\Gamma$
\citep[see Figure~4 in][]{dau13}.} A very large value of the emissivity index
will produce two effects. Firstly, it increases the blurring of the reflection
spectrum, broadening the Fe K line. Secondly, it shifts photons to lower
energies.  Broadening of the Fe K line is likely to be required, but the
extreme value of $q$ redshifts the line outside of the observed range. The
model then compensates by increasing the inclination, which shifts the line
back to higher energies. We notice that a similar effect of the lamppost model
bringing the inclination to more reasonable levels was previously reported by
\cite{tom14} in their analysis of Cyg~X-1 data. The large inclination of
Model~1 is interpreted as the result of this tension between the model
parameters.  We therefore disregard Model~1 and focus only on the application
of the lamppost model, which we then use to derive all the physical parameters
for the black hole system \j1752.

The results of the fits with all models (Table~\ref{tab:fits}) indicate that
the primary source of X-ray photons is located very close to the black hole,
specifically at $h \lesssim 2~R_\mathrm{Hor}$, where
$R_\mathrm{Hor}=1.063$~$R_g$ is the event horizon radius for a black hole
rotating at the Thorne limit. This again supports the idea of extreme
illumination of the inner regions of the accretion disk, causing a
relativistically broadened reflection spectrum. For these parameters, we
estimate that $\gtrsim 27\%$ of the photons emitted by the primary source will
fall into the black hole without reaching the accretion disk.   Furthermore,
the lamppost model has the capability of predicting a reflection fraction by
assuming the point-source lamppost emits isotropically in its rest frame.  The
corresponding reflection fraction for its height is $R_\mathrm{f} \gtrsim 5$,
whereas the best fits with Models~2.A and 2.B find \mIIArrf\ and \mIIBrrf,
respectively. This large difference is reconciled with the inclusion of
Comptonization with Model~3. Similar results were found between these classes
of models in the case of \gx339 \citep{gar15,ste17}. This difference is
because the coronal scattering dilutes reflection's apparent strength
\citep[e.g.;][]{ste16}, so that larger $R_\mathrm{f}$ is required to fit the
data.

Figures~\ref{fig:plots-1} and \ref{fig:plots-2} show that the correlations
between coronal height, inner disk radius and inclination in Model~3.A are
mostly consistent with those of Model~2.A, aside from the differences described
above for $\Gamma$ and $kT_e$. For Models~3, assuming a uniform-density corona,
its optical depth is given by $f_\mathrm{sc}=1-\exp(-\tau)$, and we
correspondingly find $\tau=1.8^{+0.1}_{-0.5}$ and $\tau=1.7^{+0.2}_{-0.4}$ for
Models~3.A and 3.B, respectively. Comparison with \citet{tit94} shows that this
temperature and optical depth are well-matched to the fitted $\Gamma$ under the
assumption of a compact {\em spherical} geometry for the corona.  Models~3.A
and 3.B give statistically indistinguishable values of the inner disk radius
and spin respectively compared to Models~2.A and 2.B, but a significant
improvement in the fit quality ($\Delta\chi^2/\Delta\nu \approx 26/1$).

\begin{figure}[h]
\centering
\includegraphics[width=\linewidth]{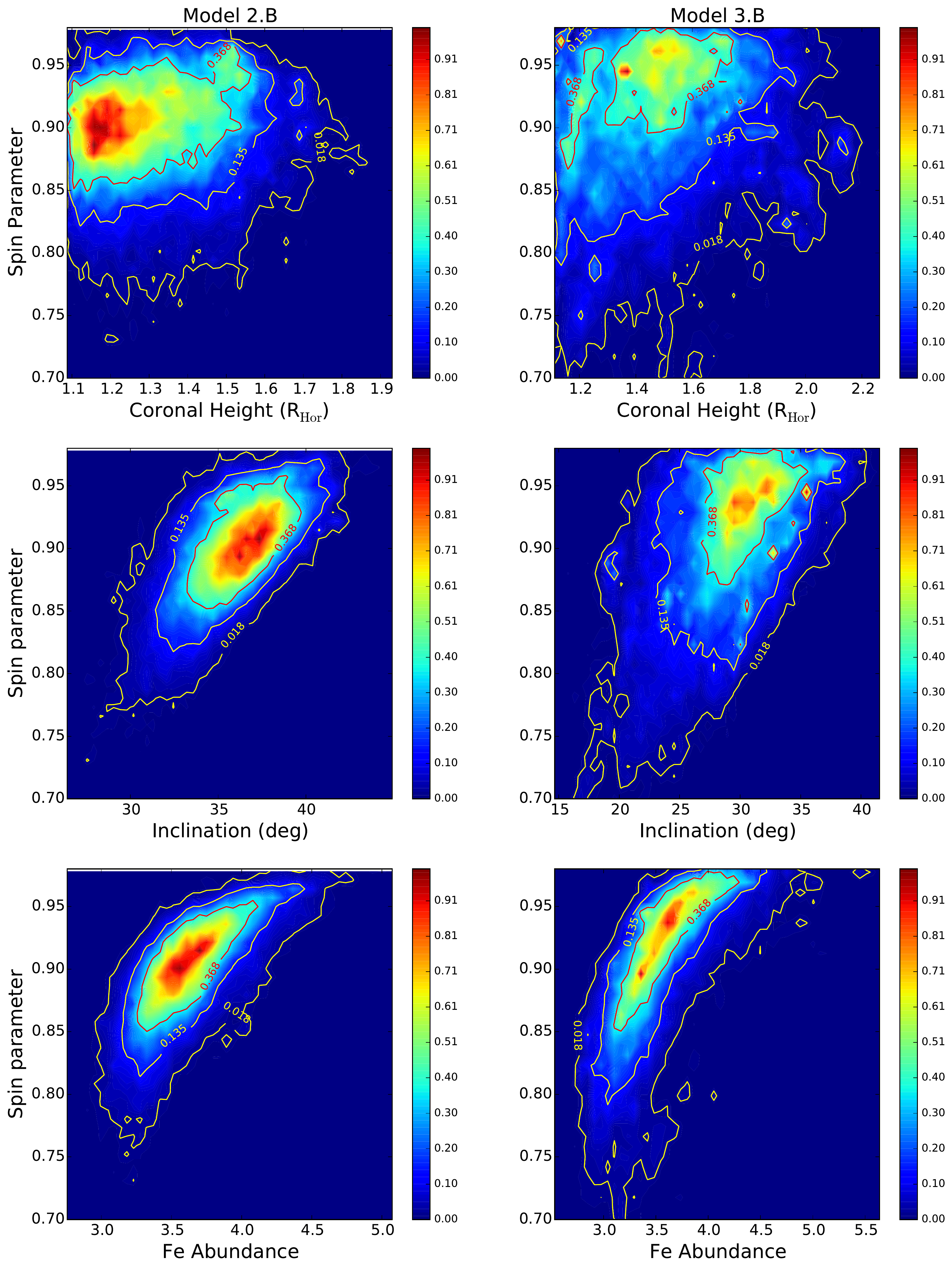}
\caption{Contour plots for the spin parameter as a function of the coronal
height, the inclination, and the Fe abundance, as obtained from the MCMC
analysis with Model~2.B and Model~3.B.
}
\label{fig:plots-spin}
\end{figure}

In all model variants we find that the inner-disk must be very close to the ISCO, leaving
vanishing room for the possibility of disk truncation. Accordingly, the black
hole spin must be quite high. The fits (Model~2.B: $a_*=$\mIIBa, Model~3.B:
$a_*=$\mIIIBa) are actually lower limits, as the presence of any truncation
effect would result in a higher spin. Figure~\ref{fig:plots-spin} shows the
contour maps for the spin in Models~2.B and 3.B versus three other important
parameters: the coronal height, the inclination, and the iron abundance. While
some correlations can be seen among these parameters, it is clear that under the
assumption that the disk reaches the ISCO, spin is very well constrained.
Perhaps the most obvious degeneracy is observed between the spin and the Fe
abundance, which is a correlation expected and previously reported in fitting
reflection models \citep[e.g.;][]{rey12,gar15}.
%

%
\graphicspath{{../plots/}}

The high spin value obtained with Model~2.B and Model~3.B disagrees with that
from \cite{rei11}, who have previously reported an intermediate spin of
$a_*=0.52\pm0.11$, the only other determination obtained through reflection
spectroscopy. We now discuss some of the possible reasons for this discrepancy.
The two observations analyzed by \cite{rei11} were taken during the decay of
the outburst, one in the intermediate state (with \suzaku) and another in the
low hard state (with \xmm). The data were modeled with the reflection model
{\tt REFBHB} \citep{ros07}.  Although this model self-consistently includes the
thermal disk emission, it also assumes a single-temperature accretion disk and
implements outdated atomic data. More importantly, the iron abundance is fixed
at the solar value. As described Section~6.3 of \cite{gar15} \citep[as well as
in][]{wan18}, the Fe K emission profile from a truncated disk with Solar
abundances looks very similar to that from a disk that extends down to the ISCO
but for which the Fe abundance is enhanced. These two situations can only be
clearly differentiated in the $\sim$10--20~keV range \citep[see Figure~12
in][]{gar15}, which is coincidentally the region not covered by the data
analyzed in \cite{rei11}.

Another important aspect of the analysis presented by \cite{rei11} is that due
to the high count rate of the source, the \xmm\ observation was taken in the
timing mode. We have found, through the analysis of \j1752\ and other sources,
serious discrepancies between the \xmm\ data taken in this mode and
simultaneous data taken by other instruments such as the PCA in \rxte.  These
discrepancies are likely due to calibration uncertainties in the \xmm\ timing
mode This discrepancy has also been noticed in the observations of the bright
hard state of \gx339 \citep{bas15}. Despite the fact that \cite{rei11} also
included one \suzaku\ observation in their work, it is likely that the \xmm\
data is dominating the results of their fits, since the count rate of the \xmm\
spectrum is between 1 and 2 orders of magnitude larger in the Fe K band (e.g.;
see their Figure~6).

In general, the results from our fits to \j1752\ resemble those found
previously for \gx339 in its bright hard state \citep{gar15,ste17,wan18}. We
find an accretion disk approaching very close to the ISCO, with a large Fe
abundance with respect to the Solar value, a rapidly rotating black hole,
and strong supporting evidence for the importance of accounting for
Comptonization of the reflection emission.
%


\section{Summary and Conclusions}\label{sec:concl}

We have presented an analysis of the bright hard state of the black hole binary
system \j1752 during its 2009 outburst observed by \rxte. During the rise of
the outburst this source was observed in a particularly stable hard state
lasting roughly one month. By combining observations taken during this period of
time, we have been able to obtain an spectrum with remarkable statistical
weight: a total of $\sim 100$~million source counts between PCU-2 and HEXTE
bands. We find the \relxill\ models are very successful in describing these
data through a combination of a thermal Comptonization continuum, unblurred
distant reflection, and relativistically blurred reflection components.
Despite the extreme statistical quality of this dataset, and the low systematics
included---0.1\% per the use of our \pcacorr\ tool---the fit statistics are
satisfactory ($\chi_{\nu}^2\sim 1.3-1.6$). This is, to our knowledge, the
highest signal-to-noise X-ray reflection spectrum published to date.

We found that the data can be almost equally well described with either an extended
coronal model or a lamppost geometry. In the former, the required emissivity
index is extreme, which suggests a compact emitting region for the primary
source of photons. When the lamppost geometry is implemented, all parameters
are consistent with the coronal model with the exception of the inclination and
reflection fraction. Inclination changes most dramatically, from $\sim67$~deg
for the coronal model to $\sim30-35$~deg for the lamppost models. The lamppost
results agrees with the upper limit of $i<49$~deg reported by \cite{mil11} from
radio jet observations, and it is thus preferred.

The modeling of the reflection spectrum of \j1752 shares several similarities
with the parameters found previously for the bright hard-state in \gx339: a
rapidly rotating black hole with an accretion disk extending very close-in with
super-solar iron abundance. Likewise, without accounting for Comptonization of
the reflection emission, the reflection fraction is found to be much lower than
that predicted self-consistently by the lamppost geometry. While the high
spin result contradicts the intermediate spin values derived by \cite{rei11},
we argue that this discrepancy is likely due to calibration uncertainties in
their data, or possibly due to the lack of coverage in the
10--30~keV region, a spectral band that is crucial to disentangle high from low
spin models, particularly if Solar Fe abundance is assumed.

Just as in the case of \gx339, we found that the reflection spectrum of \j1752
is largely affected by relativistic smearing, which requires  the inner
accretion disk to be located very close to the black hole. This result would
then suggest that at the bright-end of the hard state, the accretion disk in
these systems approaches close to, or reaches the ISCO, before their transition
to the soft state. The fact that \j1752\ spent an entire month in the bright
hard state, without showing any significant changes in luminosity or spectral
hardness, suggests that the system must have been in a very stable
configuration, which is plausible if the disk has reached the ISCO.  If this
interpretation is correct, it could possibly mean that the transition to the
soft state is then triggered not by a change in the disk's geometry, but rather
by a different physical process. A sudden change in the accretion rate could
presumably induce such a change, increasing the temperature of the accretion
disk and pushing the system into the soft state. The present discussion is
merely speculative at this point, which is why intense monitoring campaigns of
this and similar sources are highly motivated to better understand their
physical nature.
%

%
%
%
\acknowledgments The authors dedicate the present work to the memory of our
friend, colleague, and mentor Jeffrey E. McClintock, who passed on November 8,
2017. This paper grew from a program to explore black hole systems using
\rxte\ under Jeff's leadership. We thank Jeff for his mentorship, his always
open door, and for his friendship. Jeff's love for science inspired us, his
relentless curiosity was a joy to share, and we miss him greatly.

This work was partially supported under NASA contract No. NNG08FD60C.
JAG and RMTC acknowledge support from NASA grant 80NSSC177K0515. JAG also
acknowledges support from the Alexander von Humboldt Foundation. JFS has been
supported by NASA Einstein Fellowship grant PF5-160144.

{\it Facility: RXTE}
%
%
%
%
\bibliographystyle{apj}
\bibliography{my-references}
%
%
%
%
\end{document}